\documentclass[runningheads]{llncs}
\usepackage{xspace}
\usepackage{enumerate}
\usepackage{lipsum}
\usepackage{multirow}
\usepackage[shortlabels]{enumitem}
\usepackage{amssymb}
\usepackage{makecell}
\usepackage{booktabs} 
\usepackage{amsmath}
\usepackage{mathbbol}
\usepackage{float}
\usepackage{amsmath}
\usepackage{graphicx}
\usepackage{tabularx}
\usepackage{hyperref}
\usepackage{xcolor}
\usepackage{caption}
\usepackage{subcaption}
\usepackage{footnote}
\usepackage{hyphenat}
\usepackage{cite}
\newcommand{\sys}{\texttt{slytHErin}}
\newcommand{\descr}[1]{\smallskip \noindent \textbf{#1}}
\newcommand{\descrit}[1]{\smallskip \noindent}
\newcommand{\comment}[1]{}
\begin{document}

\title{\sys{}: An Agile Framework for Encrypted Deep Neural Network Inference}
\titlerunning{\sys{}: An Agile Framework for Encrypted DNN Inference}
%
\author{Francesco Intoci*\inst{1} \and
Sinem Sav*\inst{1} \and
Apostolos Pyrgelis\inst{1} \and Jean-Philippe Bossuat\inst{2} \and Juan Ramón Troncoso-Pastoriza\inst{2} \and Jean-Pierre Hubaux\inst{1,2}
}
\authorrunning{F. Intoci et al.}
%
\institute{EPFL, Lausanne 1015, Switzerland \\ \email{name.surname@epfl.ch}\\ \and
Tune Insight SA, Lausanne 1015, Switzerland\\ 
\email{name@tuneinsight.com}}
%
\maketitle              
\def\thefootnote{*}\footnotetext{These authors contributed equally to this work.}\def\thefootnote{\arabic{footnote}}

\begin{abstract}
Homomorphic encryption (HE), which allows computations on encrypted data, is an enabling technology for confidential cloud computing. One notable example is privacy-preserving Prediction-as-a-Service (PaaS), where machine-learning predictions are computed on encrypted data. However, developing HE-based solutions for encrypted PaaS is a tedious task which requires a careful design that predominantly depends on the deployment scenario and on leveraging the characteristics of modern HE schemes. Prior works on privacy-preserving PaaS focus solely on protecting the confidentiality of the client data uploaded to a remote model provider, e.g., a cloud offering a prediction API, and assume (or take advantage of the fact) that the model is held in plaintext. Furthermore, their aim is to either minimize the latency of the service by processing one sample at a time, or to maximize the number of samples processed per second, while processing a fixed (large) number of samples. In this work, we present \sys{}, an agile framework that enables privacy-preserving PaaS beyond the application scenarios considered in prior works. Thanks to its hybrid design leveraging HE and its multiparty variant (MHE), \sys{} enables novel PaaS scenarios by encrypting the data, the model or both. Moreover, \sys{} features a flexible input data packing approach that allows processing a batch of an arbitrary number of samples, and several computation optimizations that are model-and-setting-agnostic. \sys{} is implemented in Go and it allows end-users to perform encrypted PaaS on custom deep learning models comprising fully-connected, convolutional, and pooling layers, in a few lines of code and without having to worry about the cumbersome implementation and optimization concerns inherent to HE. 

\end{abstract}
\section{Introduction}\label{sec:intro}

With recent advances in deep learning, cloud service providers expose trained deep neural networks (DNNs) to end-users for prediction-as-a-service (PaaS) through their application programming interfaces (APIs)~\cite{awsml,googleml,azure,bigml,watson}. For instance, Amazon Forecast enables business analytics by performing forecasting on client time-series data~\cite{amazonForecast} and Azure's Cognitive summarizes and classifies financial documents~\cite{azurecognitive}. However, PaaS applications raise privacy concerns as both the user data (e.g., client time-series, text, or health data) and the machine learning model (due to intellectual property concerns), can be sensitive information, and cloud service providers must comply with privacy regulations such as CCPA~\cite{CCPA}, GDPR~\cite{GDPR}, and HIPAA~\cite{HIPAA}. Thus, it is now needed more than ever to protect the privacy of the data used in PaaS applications.

To enable privacy-preserving PaaS, various works propose performing encrypted DNN inference by employing homomorphic encryption (HE) schemes which allow computations directly on ciphertexts~\cite{CryptoNets,fasterCN,MiniONN,Gazelle,riazi2019xonn,mishra2020,cheetah2022,boemer2019ngraph,cryptoDL,brutzkus2019low,zama,doren,guillermo,Boura2018ChimeraAU}. However, to cope with the computational overhead introduced by HE operations and to account for the characteristics of modern HE schemes, e.g., their support of Single Instruction, Multiple Data (SIMD) operations, these works rely on various optimizations which are tailored to specific PaaS scenarios, the most common of which comprises a cleartext DNN model and encrypted data. As a result, existing HE-based works cannot support emerging scenarios, e.g., edge machine learning~\cite{10.1145/3469029, 10.1145/3510831, 10069495}, that require outsourcing the prediction to the client (while protecting the model's intellectual property), or privacy-preserving federated learning where inference is performed on a model that is trained in encrypted form by multiple data providers~\cite{poseidon, rhode, 9935302}. Moreover, these works rely on data packing schemes adapted to specific DNN architectures and application requirements, aiming either to minimize prediction latency (typically by processing one sample at a time, e.g., for real-time analytics), or to maximize the number of samples processed per second (usually by collecting and then processing in parallel a large number of samples leveraging on SIMD capabilities).



In this work, we design \sys{}, an agile framework for encrypted DNN inference. Built on HE and its multiparty variant, our framework can be adapted to various and novel PaaS scenarios where: (i) the client's data is encrypted while the model is in cleartext, (ii) the client's data is in cleartext and the model is encrypted, and (iii) both the client's data and the model are encrypted. 
Moreover, \sys{} features application- and model-agnostic optimizations which make it suitable for various settings. For instance, \sys{} implements an intuitive and flexible packing scheme that efficiently enables SIMD operations for \emph{arbitrary} batch sizes, and generic optimizations for encrypted matrix operations. We implement \sys{} in Go and provide the building blocks that enable the encrypted execution of any DNN model composed of fully-connected, convolutional, and pooling layers. Contrary to prior works, our implementation is not centered around a system model, specific assumptions, or DNN architectures, making it a versatile tool for securing different PaaS pipelines. Our evaluation shows that \sys{} achieves accuracy similar to performing inference on cleartext data and/or models. Moreover, it yields an interesting trade-off between latency and throughput, and its overall performance is on par with that of the state-of-the-art HE-based inference solutions, while being more flexible than specialized solutions. Our implementation can be found on \url{https://github.com/ldsec/slytHErin}.
\section{Related Work}\label{sec:related}


Given the potential privacy issues that might arise in PaaS, a number of works that build encrypted PaaS frameworks have been proposed. These works rely on homomorphic encryption (HE) and/or multiparty computation (MPC) to protect the confidentiality of both the ML model and the client's evaluation data during prediction~\cite{CryptoNets,MiniONN,Gazelle,blaze,riazi2019xonn,mishra2020,cheetah2022,boemer2019ngraph,cryptoDL,Deevashwer2020,brutzkus2019low,zama,doren,guillermo,Boura2018ChimeraAU,Chillotti2020}.

\descr{HE-based Solutions.} Cryptonets was the first work in this research direction that enabled DNN evaluation on encrypted data using an HE scheme~\cite{CryptoNets}. Its overhead, in terms of latency, was later improved by Brutzkus et al. which proposed novel approaches to represent the input data~\cite{brutzkus2019low}. Other works focus on improving the efficiency of encrypted matrix operations~\cite{10.1145/3267973.3267976,Jiang2018} or on designing novel techniques for  the encrypted evaluation of more complex ML models such as graph convolutional networks~\cite{ran2022cryptogcn}. The latter has been used in downstream tasks such as human action recognition~\cite{Kim2022} achieving better latency than~\cite{brutzkus2019low}. Other works develop compilers that ease the deployment of trained ML models with HE libraries, e.g., SEAL~\cite{seal}, HElib~\cite{helib}, or Palisade~\cite{palisade}, for encrypted inference. Boemer et al.~\cite{boemer2019ngraph,boemer2018ngraph} build a graph compiler for SEAL that simplifies the use of a model trained with Tensorflow~\cite{tensorflow} or PyTorch~\cite{pytorch} for encrypted PaaS. CHET, on the other hand, is a domain-specific optimizing compiler that allows the specification of tensor circuits suitable for HE-based DNN inference~\cite{CHET}. All of these works propose specific input data representations (packing) and optimizations for either latency or throughput for specific scenarios (e.g., featuring a cleartext model vs. encrypted data) and DNN architectures. Moreover, to cope with DNN non-linear operations that are not supported by HE schemes, e.g., activations, they either use interactions with the client~\cite{boemer2019ngraph}, modify their functionality to low-degree polynomial functions~\cite{CryptoNets,brutzkus2019low,Kim2022,CHET}, or use polynomial approximations~\cite{ran2022cryptogcn,Chabanne2017PrivacyPreservingCO}.

\descr{Hybrid Approaches.} To ease the encrypted execution of non-linear functions, some works rely on hybrid approaches combining two-party computation with HE~\cite{Gazelle,MiniONN,cheetah2022,Deevashwer2020}, or secret sharing with garbled circuits~\cite{blaze,riazi2019xonn,mishra2020}. For instance, Liu et al.\cite{MiniONN} utilize HE for matrix multiplications and garbled circuits for the non-linear activations. Juvekar et al.~\cite{Gazelle} employ HE for matrix-vector multiplication and convolution operations and garbled circuits for comparisons which are widely used in activation functions. Similarly, we provide a hybrid framework for privacy-preserving PaaS that supports a wide range of applications by relying on a multiparty variant of HE. Moreover, thanks to our generic data representation scheme and optimizations, our framework is agnostic of the DNN architecture and parameters such as batch size, while achieving on par performance with the state-of-the-art.



\section{Background}\label{sec:background}
\subsection{Homomorphic Encryption}\label{sec:mhe}

Homomorphic encryption (HE) schemes enable the execution of arithmetic operations directly on ciphertexts, i.e., without requiring decryption; this makes them ideal candidates for privacy-preserving machine learning inference applications. In this work, we employ the Cheon-Kim-Kim-Song (CKKS) scheme~\cite{cheon2017homomorphic}, which is suitable for machine learning tasks as it enables approximate arithmetic over $\mathbb{C}^{\mathcal{N}/2}$ (hence, over real values as well). The ring $R_{Q_{L}}=\mathbb{Z}_{Q_{L}}[X]/(X^\mathcal{N}+1)$ of dimension $\mathcal{N}$ with coefficients modulo $Q_{L}=\prod_{i=0}^{L}q_{i}$ defines the plaintext and ciphertext spaces, hence both plaintexts/ciphertexts are represented by polynomials of degree $\mathcal{N}-1$ whose coefficients encode a vector of $\mathcal{N}/2$ values. The security of CKKS is based on the ring learning with errors problem~\cite{lyubashevsky2013ideal}. CKKS supports the homomorphic evaluation of operations such as additions, multiplications, and rotations, and any operation is simultaneously performed on all encoded values, hence offering Single Instruction, Multiple Data (SIMD). Non-linear operations, e.g., comparisons, are supported via polynomial approximations, introducing a computation overhead versus accuracy tradeoff. CKKS is a leveled HE scheme, i.e., an \textit{L}-depth circuit can be evaluated before the ciphertext is exhausted. Then, a costly procedure, called bootstrapping~\cite{gentry2009}, is required to refresh the exhausted ciphertext and enable more operations on it. We refer to the traditional bootstrapping operation (performed by a single party) as \emph{centralized bootstrapping}.

\descr{Multiparty Homomorphic Encryption (MHE).} To make our framework adaptable to various PaaS scenarios (see Section~\ref{sec:design}), we also rely on a multiparty variant of the CKKS scheme~\cite{mouchet2019distributedbfv}. In the multiparty homomorphic encryption (MHE) scheme, a set of parties (e.g., model-providers) collectively generate a public key while the corresponding secret key is secret-shared among them. This setting enables secure collaboration between $N$ parties, as parties use the collective public key to encrypt their inputs and perform joint operations on them using the MHE scheme. The result decryption by the client, however, requires the participation of all parties. Hence, this scheme ensures confidentiality under a passive adversary model with up to $N-1$ collusions. Moreover, the multiparty CKKS scheme offers efficient multiparty computation protocols. For instance, it enables a collective bootstrapping operation, where the costly centralized bootstrapping which homomorphically evaluates the decryption and consumes many levels, is substituted by a lightweight one-round interactive protocol ($\textsf{CBootstrap}(\cdot)$) which does not consume levels. Moreover, the scheme supports ($\textsf{CKeySwitch}(\cdot)$), a collective key-switch operation which can change the encryption key of a ciphertext.

\subsection{Deep Neural Networks}

Deep neural networks (DNNs) are able to model complex non-linear relationships and find applicability in various domains such as computer vision. A DNN consists of multiple hidden layers between the input and output layers. Our framework enables the encrypted evaluation of DNNs comprising fully connected (\textsc{FC}), convolutional (\textsc{Conv}), and pooling (\textsc{Pool}) layers. We succinctly present the functionalities of these layer types:

\descr{Fully Connected layer}: Given an input vector $\mathbf{x}$, a weight matrix $\mathbf{W}$ and a bias vector $\mathbf{b}$, a \textsc{FC}-layer computes $\mathbf{x}\mathbf{W^T}+\mathbf{b}$.

\descr{Convolutional layer}: Given an input tensor (e.g., an image) $\mathbf{X}$ with $c_i$ channels of dimensions $w \cdot h$ and a set of $c_o$ kernels $\mathbf{K}$ each made up of $c_i$ filters of size $f_w\cdot f_h$, a \textsc{Conv}-layer computes a tensor $\mathbf{O}$ with $c_o$ channels. Each channel $\mathbf{O}_i$ is computed as $\sum_{n=0}^{c_i} \mathbf{X}_n * \mathbf{K}_{i,n}$, with $\mathbf{X}_n$ the n-th channel of the input image, $\mathbf{K}_{i,n}$ the n-th filter of the i-th kernel, and $*$ the cross-correlation operator. 

\descr{Pooling layer}: It performs dimensionality reduction on the input. The most common types are \textsf{SumPooling}, \textsf{AveragePooling}, and \textsf{MaxPooling}, where the feature-map is the sum, average, and the maximum of the features in a region of the input, respectively. \textsf{Max-Pooling} requires non-linear operations, i.e., comparisons, which are non-trivial to implement under encryption, thus we only consider the first two types.

Each layer can be paired with an activation function which is evaluated on its output. The output of the DNN's last layer is the prediction result (output).



\section{\sys{} Overview}\label{sec:design}

\begin{figure*}[t]
\centering
    \centering
    \includegraphics[width=\textwidth]{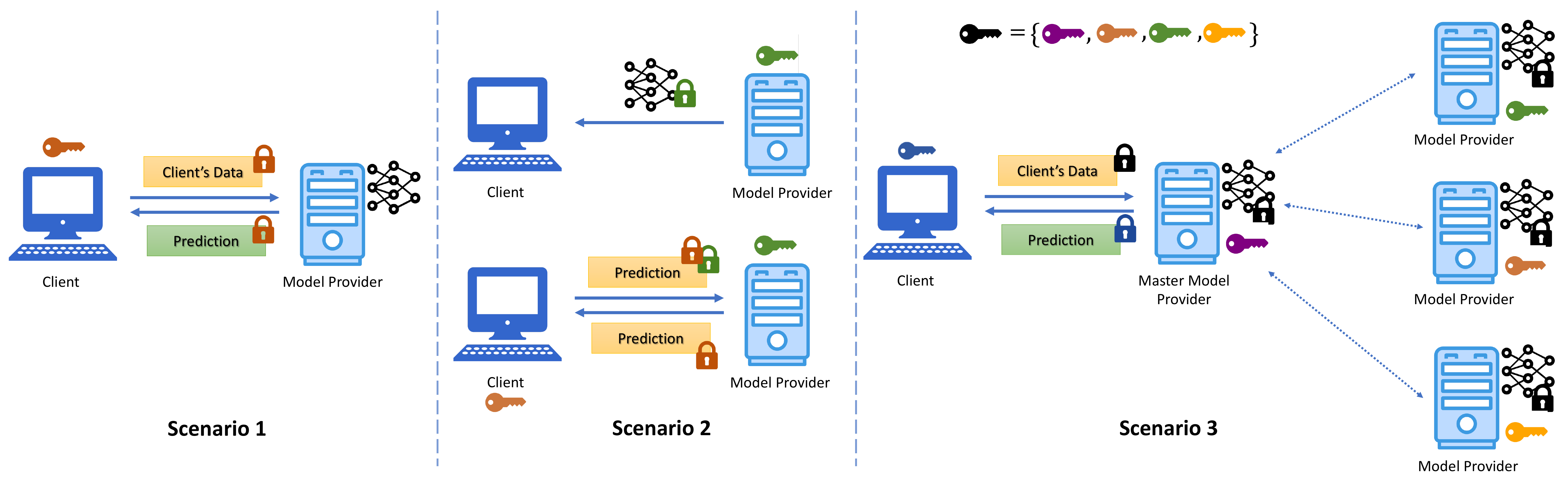} 
    \caption{Encrypted PaaS scenarios enabled by \sys{}. Encryption is depicted with a lock whose color is the same as the corresponding secret key. The black key (rightmost figure) corresponds to the model-providers' collective key. \textbf{Scenario 1:} The client sends its encrypted data to the model-provider that evaluates it on the plaintext model. \textbf{Scenario 2:} The encrypted model is sent to the client for evaluation on its cleartext data. \textbf{Scenario 3:} The client sends encrypted data to a cohort of model-providers that retain an encrypted model.}
    \label{fig:scenarios}
\end{figure*}

Building on the CKKS HE scheme and its multiparty variant (see Section~\ref{sec:background}), we design a framework that is flexible for various encrypted PaaS scenarios (Figure~\ref{fig:scenarios}). We first describe the involved entities before detailing \sys{}'s objectives and workflow for each PaaS scenario.


\begin{itemize}
    \item \descr{Model-provider(s):} This entity (one or more) has trained an ML model and exposes it to end-users for queries (PaaS) through a prediction API hosted on a cloud service provider.
    \item \descr{Client:} This entity is a user of the PaaS that inputs its own sensitive data which is evaluated on the model exposed by the model-provider. The client obtains the output of the PaaS process, i.e., the prediction.
\end{itemize}

We consider that the client and the model-provider are \textit{honest-but-curious}, i.e., they follow the protocol specification, but they might try to infer information about each other's data. \sys{}'s objective is to protect both the confidentiality of the client's and the model-provider's data. In particular, the model-provider should not learn any information about the client's evaluation data and the prediction result, whereas the client should not obtain any knowledge about the model beyond what can be inferred from the PaaS output.

\subsection{Scenario 1: Encrypted client data - Cleartext model}
\label{sub:scenario1}

This is the traditional HE-based PaaS setting, where a client encrypts its data with its own public key and sends the ciphertext to the model-provider that stores its ML model in plaintext form. The model-provider evaluates its model on the client's encrypted data -- without interacting with the client -- and returns the encrypted prediction to the client. The client decrypts the ciphertext with its secret key and obtains the prediction result. In this scenario, the client's data confidentiality is ensured as its inputs are encrypted throughout the DNN evaluation and the model-provider does not learn the prediction result. The model confidentiality is protected as the model remains on the model-provider's side. Scenario 1 represents a typical PaaS setting, where a model-holder exposes a prediction service that receives sensitive data as inputs~\cite{CryptoNets,zama,brutzkus2019low,Gazelle}. For instance, imagine a health-care insurance provider that uses its customer data and trains a DNN that predicts the probability of patient re-admission to a hospital. The model is exposed through an API to clients (e.g., hospitals) who wish to obtain predictions about their own cohorts of patients. However, hospitals cannot share their patient data with third-parties due to ethical and data privacy requirements, hence, \sys{} could be an enabler for such a service as it ensures data confidentiality.

\subsection{Scenario 2: Cleartext client data - Encrypted model}
\label{sub:scenario2}

In this scenario, the model-provider outsources the computation of the prediction to the client. However, the model is an intellectual property that needs to be protected. Thus, the model-provider encrypts its model with its own public key and sends it to the client in encrypted form. The client evaluates the encrypted model on its own (plaintext) data and obtains an encrypted prediction. Finally, the client sends the prediction ciphertext to the model provider, which obliviously decrypts the result and communicates it back to the client (Section~\ref{sub:mpc}). The client's data confidentiality is ensured as its evaluation data is never transferred and the model-provider does not learn the prediction result due to the oblivious decryption phase. The model confidentiality is protected as the model is encrypted with the model-provider's public key. Scenario 2 is suitable for applications that require outsourcing a trained model to the client side for predictions. For instance, this could be the case for model trading platforms that offer a \emph{try-before-you-buy} option, where customers locally test the performance of an ML model on their data before purchasing it. Another relevant application is model outsourcing to edge devices~\cite{10.1145/3469029, 10.1145/3510831}, e.g., mobile phones or smartwatches, that monitor their owners' activity and provide feedback to them through predictions, e.g., health recommendations or activity tracking~\cite{10069495}. We note that this is a novel PaaS scenario enabled by \sys{}.

\subsection{Scenario 3: Encrypted client data - Encrypted model}
\label{sub:scenario3}
In this scenario, we assume that the model-provider is represented by a cohort of $N$ nodes that have collectively trained a DNN on their joint data with a state-of-the-art encrypted collaborative learning framework~\cite{poseidon,rhode,9935302}. For this, we rely on a multiparty variant of homomorphic encryption (MHE). In particular, the nodes (model-providers) generate a collective public key (black key in Figure~\ref{fig:scenarios}, \textbf{Scenario 3}) whose corresponding secret key is secret-shared among them (colored keys in Figure~\ref{fig:scenarios}, \textbf{Scenario 3}). We assume that the nodes collectively train a DNN on their data and retain it under encryption for PaaS to mitigate model-targeting attacks and protect its intellectual property. For this scenario, the client encrypts its evaluation data with the collective public key and a master node from the cohort performs the prediction (with both the model and the data encrypted) with the assistance of the other nodes for collective interactive operations (e.g., ciphertext refresh -- $\textsf{CBootstrap}(\cdot)$, Sections~\ref{sec:mhe} and~\ref{sub:mpc}). Finally, the ciphertext storing the prediction result is re-encrypted (i.e., $\textsf{CKeySwitch}(\cdot)$, Sections~\ref{sec:mhe} and~\ref{sub:mpc}) under the public key of the client which decrypts it to obtain the prediction. In this case, both the model and the client data are encrypted with the cohort's collective public key, hence, their confidentiality is ensured as long as one of the cohort nodes is honest and does not participate in decryption. The confidentiality of the prediction output is protected, as only the client can decrypt it. Scenario 3 is suitable for PaaS applications after a model-provider outsources the model training procedure to a cohort of $N$ nodes that leverage on distributed learning techniques for improved efficiency or after a federation of $N$ model-providers, each with their own data, uses a state-of-the-art framework to train a collective ML model \emph{under encryption}~\cite{spindle,poseidon,rhode}. We note that previous works that focused on encrypted DNN inference do not support (or implement) inference on encrypted models or collaborative functionalities such as bootstrapping or re-encryption.

\section{Cryptographic Building Blocks}\label{sec:crypto}
We describe \sys{}'s underlying cryptographic building blocks that make it flexible and efficient for different encrypted PaaS scenarios (Section~\ref{sec:design}) and various DNN architectures. We first introduce the data packing approach adopted to encode/encrypt the input data (Section~\ref{sub:packing}). Then, we describe the algorithms used to evaluate fully-connected, convolutional, and pooling layers under encryption in Sections~\ref{sub:multiplication} and~\ref{sub:convolutions}, respectively. We also present several optimizations that \sys{} implements (Section~\ref{sub:optimizations}) and how non-linear activation functions are evaluated (Section~\ref{sub:nonlinear}). Finally, we present the multiparty computation protocols which allow \sys{} to support novel PaaS scenarios (Section~\ref{sub:mpc}).


\subsection{Input Data Packing}\label{sub:packing}


Modern homomorphic encryption schemes can encode (pack) a vector of values into one ciphertext, thus enabling SIMD operations via the parallel computation of a function on all ciphertext slots. Designing an efficient packing scheme is crucial, yet challenging, due to the costs of re-arranging the ciphertext slots via rotations. Prior work on encrypted DNN inference~\cite{brutzkus2019low,CryptoNets,Gazelle,MiniONN,Kim2022,Deevashwer2020,cheetah2022} designed efficient packing schemes but these are tailored to specific system models and assumptions (e.g., the client's availability for the evaluation of certain operations). \sys{} employs a simple yet generic data packing scheme that is agnostic of the encrypted PaaS scenario and also flexible in terms of batch size that results in optimized latency and throughput.
Given a batch consisting of $n$ input samples each with $d$ features, a naive approach is to encrypt/encode each feature of an input sample separately, yielding an inefficient execution due to the high number of ciphertexts/plaintexts. To leverage on SIMD operations and enable efficient encrypted inference, we flatten the batch and encrypt/encode all values in a single ciphertext/plaintext. For an input sample represented by a tensor of size $h \times r \times c$ (where, e.g., for an image, $h$ is the number of channels, while $r$ and $c$ represent the size of the pixel matrix of each channel), we encrypt/encode a batch of size $n$ in a tensor of size $n \times h \times r \times c$ as follows: First, we row-flatten (\textsf{RowFlatten($\cdot$)}) each of the $n$ tensors, such that the batch-tensor is transformed into a matrix of size $n \times d$, with $d=h \times r \times c$. This is done by iterating through all the channels of the input, by row-flattening the corresponding 2D matrix, and by horizontally stacking their flattened representation. The $n \times d$ matrix is then transposed and row-flattened (\textsf{TensorFlatten($\cdot$)}), thus yielding a vector of size $m = d \times n$. Our packing scheme requires that $m \leq s$, where $s$ is the ciphertext capacity (i.e., $s = \mathcal{N}/2$ for CKKS) and if that is not possible, we employ block matrix arithmetic optimizations (see Section~\ref{sub:optimizations}).

\subsection{Matrix Multiplication}\label{sub:multiplication}
\begin{figure}[t]
    \resizebox{1.0\linewidth}{!}{\includegraphics{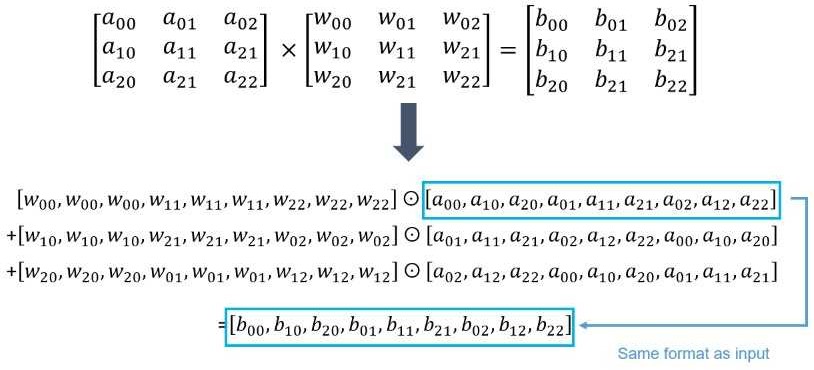}}
    \caption{Multiplication of two matrices $\mathbf{A}$ and $\mathbf{W}$ of size $3 \times 3$.}
    \label{fig:mult}
\end{figure}

To support the evaluation of fully-connected layers under encryption, \sys{} relies on the following matrix multiplication algorithm. Given two encrypted matrices, $\mathbf{A}$ and $\mathbf{W}$, where $\mathbf{A}$ is of size $n \times d$ and $\mathbf{W}$ of size $d\times h$, \sys{} implements their multiplication following the diagonal approach of~\cite{HelibArt}. First, $\mathbf{W}$ is represented by its \emph{generalized diagonals}~\cite{HelibArt}, where the element \textit{i,j} of the diagonal is: $d_{i,j} = \mathbf{W}_{(i+j) \bmod d, j}$. Additionally, we replicate $n$ times the element $d_{i,j}$. The matrix multiplication, then, can be evaluated as follows:
\vspace{-0.5em}
\[\mathbf{A} \times \mathbf{W} = \displaystyle\sum_{i=1}^{d} \mathbf{d_i} \odot \textsf{RotateCyclic}_{d \times i}(\textsf{RowFlatten}(\mathbf{A}^T))\]

\noindent where $\textsf{RotateCyclic}_{k}(\mathbf{v})$ represents a cyclic rotation of the values in $\mathbf{v}$ by $k$ positions to the left and $\odot$ represents the Hadamard product. Figure~\ref{fig:mult} represents a multiplication of two $3\times 3$ matrices with this algorithm. 


\subsection{Convolutional and Pooling Layers}\label{sub:convolutions}

To evaluate convolutional layers under encryption, \sys{} represents the convolution operation as a matrix multiplication by expressing the filter as a \textit{Toeplitz} matrix~\cite{Toeplitz2,Toeplitz}.
For ease of presentation, consider a toy-example with a convolution between 
a single-channel input $\mathbf{I} \in R^{3 \times 3}$ and a filter $\mathbf{h} \in R^{2 \times 2}$
operating on the input with unitary stride and no padding. We can compute the convolution as:
$\mathbf{O} = \textsf{TensorFlatten}(\mathbf{h} * \mathbf{I})^T = \mathbf{h'} \times \mathbf{I'}$
where $\mathbf{I'}  = \textsf{TensorFlatten}(\mathbf{I})^T$ and $\mathbf{h'}=\mathcal{T}(\mathbf{h})$ for a function $\mathcal{T}$ that returns a Toeplitz matrix~\cite{Toeplitz} as follows:
\begin{align*}
\scalebox{0.95}{
$
\mathbf{h'}=
\begin{pmatrix}
  h_{1,1} &  h_{1,2} & 0 & h_{2,1} & h_{2,4} & 0 & 0 & 0 & 0\\
  0 & h_{1,1} &  h_{1,2} & 0 & h_{2,1} & h_{2,4} 0 & 0 & 0 & 0\\
  0 & 0 & 0 & h_{1,1} &  h_{1,2} & 0 & h_{2,1} & h_{2,4} & 0 \\
  0 & 0 & 0 & 0 & h_{1,1} &  h_{1,2} & 0 & h_{2,1} & h_{2,4}\\
\end{pmatrix}
$
}
\end{align*}
Note that computing $\mathbf{O^T} = \mathbf{I'^T} \times \mathbf{h'^T}$ allows us to utilize the matrix multiplication algorithm and the input data packing protocol of Sections~\ref{sub:multiplication} and~\ref{sub:packing}, respectively. Moreover, $\mathbf{O^T}$ is a valid input to any subsequent layer in the DNN architecture, without requiring any re-packing, hence avoiding the cost of slot re-arrangement. \sys{} generalizes this method for convolutional layers with $k$ kernels, each with $m$ filters, and $n$ inputs with $m$ channels. \sys{} also supports \textsf{SumPooling} and \textsf{AveragePooling} layers: these are evaluated by treating them as convolutional layers, and employing the method previously described.
%



\subsection{Optimizations}\label{sub:optimizations}

\descr{Complex-Number Trick.} To optimize the input data packing scheme (Section~\ref{sub:packing}), \sys{} employs the complex-number trick~\cite{Sav2022.01.10.475610}: Since the CKKS plaintext space is $\mathbb{C}^{{\mathcal{N}}/{2}}$, we can leverage the imaginary part of complex numbers and pack (up to) two values in one plaintext slot. This allows us to effectively perform the multiplication and sum of two values with just one multiplication. As a toy example, let us consider the vectors: $\mathbf{a} = (a_1,\dots), \mathbf{b} = (b_1,\dots), \mathbf{c} = (c_1,\dots),$ and $\mathbf{d} = (d_1,\dots)$. To compute $\mathbf{a} \odot \mathbf{c} + \mathbf{b} \odot \mathbf{d} = (a_1c_1 + b_1d_1 , \dots)$, we compress the first two and the two last vectors each into one vector with the following complex representation:
$\mathbf{g} = (a_1 + ib_1 , \dots)$, $\mathbf{h} = (c_1 - id_1 , \dots)$.
Then, $\mathbf{g} \odot \mathbf{h} = (a_1c_1 + b_1d_1 + ie , \dots)$ for some value $e$, and the real part of the result can be extracted with complex conjugation, addition and constant multiplication.
We apply this technique to the input matrix $\mathbf{A}$ and to the weight matrix $\mathbf{W}$. In particular, we embed pairs of adjacent columns of $\mathbf{A}$ into one column, i.e., column $k$ is paired with column $k+1 \bmod d$, where $d$ is the number of columns, hence the entry $\mathbf{A}_{(k,j)}$ becomes $\mathbf{A}_{(k,j)}+i\mathbf{A}_{(k,j+1)}$. For $\mathbf{W}$, we compress the pairs of adjacent \textit{diagonals} into one, padding with an extra $0$-diagonal if the number of diagonals is odd. The newly packed matrix $\mathbf{\Tilde{W}}$ has $\lceil \frac{d}{2} \rceil$ diagonals instead of $d$, reducing the complexity of the matrix multiplication algorithm by a factor of $2$.



\begin{figure}[t]
    \resizebox{1.0\linewidth}{!}{\includegraphics{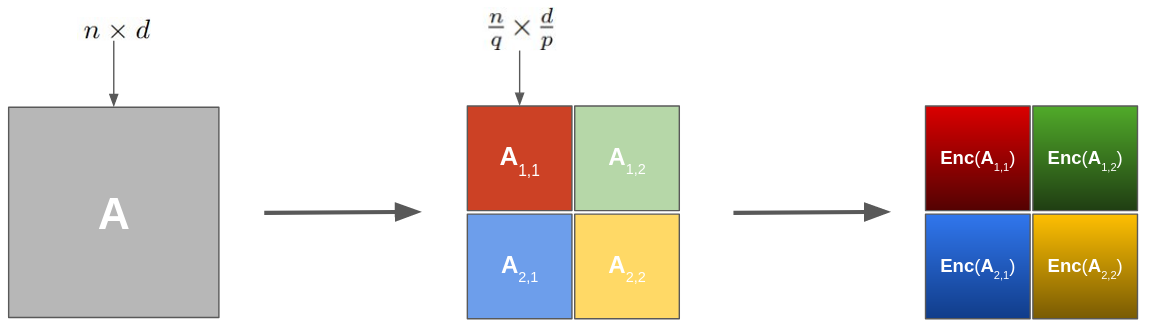}}
    \caption{Partitioning of an input matrix $\mathbf{A}$ in a $2 \times 2$ block matrix.}
    \label{fig:block}
\end{figure}

\descr{Block Matrix Arithmetic.} When the size of the input batch exceeds the ciphertext capacity, \sys{} employs block-matrix arithmetic~\cite{rhode}. The input matrix $\mathbf{A}$ of size $n \times d$, is represented as a block-matrix $\mathbf{\Bar{A}}$ of size $q \times p$, i.e., a matrix consisting of \emph{blocks} (or sub-matrices) of size $\frac{n}{q} \times \frac{d}{p}$ for some divisors $q$ and $p$ of $n$ and $d$, respectively. Similarly, the weight matrix $\mathbf{W}$ of size $d \times h$ is partitioned to enable the multiplication $\mathbf{\Bar{O}} = \mathbf{\Bar{A}} \times \mathbf{\Bar{W}}$ under two constraints: (i) $\mathbf{\Bar{W}}$ must have $p$ row partitions, and (ii) every inner block $\mathbf{W}_{k,j}$ must be compatible for matrix multiplication with the inner blocks $\mathbf{A}_{i,k}$. $\mathbf{\Bar{O}}$ is a block-matrix of size $n \times h$ with $q$ row partitions and $m$ column partitions (and $m$ the number of column partitions of $\mathbf{\Bar{W}}$). Each block $\mathbf{O}_{i,j}$ is computed as: $\mathbf{O}_{i,j} = \sum_{k=1}^{p} \mathbf{A}_{i,k}\mathbf{W}_{k,j}$.
\noindent Hence, by choosing suitable partitions, each matrix inner block is small enough to be encrypted/encoded independently following the input data packing and the generalized-diagonals approach described earlier (Sections~\ref{sub:packing} and~\ref{sub:multiplication}). Figure~\ref{fig:block} represents the encryption of matrix $\mathbf{A}$ with $2\times 2$ partitioning. Then, the matrix multiplication between two large matrices is evaluated as a series of sums and multiplications between these smaller blocks.
Given a model to evaluate (i.e., the dimensions of its layers), the number of input features, and a set of CKKS parameters, \sys{} follows a heuristic-based approach to automatically find the best batch size and partition strategy. In more detail, \sys{} explores the space of possible splits, starting from divisors of the number of samples (if provided by the user) or divisors of the features dimension, and picks the split sequence and batch size that minimize the overall complexity of the pipeline in terms of homomorphic operations (i.e., it minimizes the number of homomorphic multiplications required to evaluate the model), thus optimizing throughput. In any case, the user can also declare a customized batch size which overrides the optimized batch size, and let \sys{} operate with a sub-optimal block matrix representation. An advantage of the block matrix arithmetic approach is that it is amenable to parallelization: Given $q \times p \times m$ threads, the matrix multiplication between two blocks $\mathbf{A}_{i,k} $ and $\mathbf{W}_{k,j}$ can be delegated to each thread, while using $q \times m$ of them to combine the individual results. Moreover, for a given set of cryptographic parameters and the corresponding evaluation keys, the client does not need to regenerate the keys for the evaluation of arbitrary size matrices, which is a computationally intensive task.


\subsection{Non-Linear Operations}\label{sub:nonlinear}

As non-polynomial functions, e.g., comparisons, are not computable under HE, some works modify common activation functions (e.g., $\textsf{ReLU}$) with simple polynomial functions~\cite{CryptoNets} (e.g., $x^{2}$), or use polynomial approximations~\cite{poseidon}. \sys{} employs the second approach and relies on \textit{Chebychev interpolants} to approximate any Lipschitz continuous function on any finite real interval.

\subsection{Multiparty Computation Protocols}
\label{sub:mpc}
We remind that \sys{} relies on CKKS and its multiparty variant (MHE) which enables interactive functionalities such as $\textsf{CBootstrap}(\cdot)$ for collective bootstrapping and $\textsf{CKeySwitch}(\cdot)$ for collective key-switching. The latter enables changing the encryption key of a ciphertext. In \textbf{Scenario 3}, the model-providers rely on these functionalities to refresh the ciphertexts noise and to change the encryption key of the prediction result, so that only the client can decrypt it.

We also design and implement an oblivious decryption protocol $\textsf{ObvDec}(\cdot)$, for \textbf{Scenario 2} (Section~\ref{sub:scenario2}). In this protocol, the client masks its prediction result (encrypted under the model provider secret key) with an encryption of 0 under an ephemeral secret key, and sends the result to the model provider, which can remove one layer of encryption from the result (by invoking the decryption procedure of CKKS), without exposing the underlying plaintext. The result is finally sent to the client that unmasks it.
\section{Experimental Evaluation}\label{sec:evaluation}


\subsection{Implementation and Experimental Setup}
We implemented \sys{} in Go~\cite{Go}, using Lattigo as the cryptographic library~\cite{lattigo}. Our implementation is modular, reusable, and easy to adapt to several PaaS applications. 
Detailed documentation can be found along with our source code on  \url{https://github.com/ldsec/slytHErin}. 
We evaluate \sys{} using the following DNN architectures: 

\begin{itemize}
    \item \textbf{NN5}: A 5-layer convolutional neural network described in~\cite{CryptoNets} for which we replace the square activation function with a degree 2 Chebyshev approximation of $\textsf{Softplus}$.
    \item \textbf{NN20}: A 20-layer DNN composed of convolutional and fully connected layers described in~\cite{zama} ($\sim$754K model parameters) for which we replace the activation functions with a degree-63 approximation of $\textsf{SiLU}$ and train it with the \textsf{MSE} loss function.
    \item \textbf{NN50}: Similar to \textbf{NN20} but comprising 50 layers ($\sim$1M model parameters~\cite{zama}).
\end{itemize}

We use the \textbf{MNIST} dataset~\cite{MNIST} for encrypted image classification, as it is the de-facto benchmark dataset used in prior work for privacy-preserving inference tasks~\cite{CryptoNets,fasterCN,boemer2018ngraph,boemer2019ngraph,brutzkus2019low,Gazelle,zama}. All models were trained from scratch, achieving similar accuracy to the original works (and with minimal accuracy loss in the encrypted inference, none for \textbf{NN5}, approximately $\sim$0.13$\%$ for \textbf{NN20}, and $\sim$2$\%$ for \textbf{NN50}). The CKKS parameters are configured to achieve 128-bit security. For the multiparty interactive protocols, we deploy \sys{} on a local cluster with an average network delay of 20ms and 1Gbps bandwidth. All experiments were executed on machines running Ubuntu 22.04, with 12-core Intel Xeon E5-2680 2.5 GHz CPUs and 256GB RAM DDR4. The results are averaged over 3-5 runs.

\subsection{Empirical Results}\label{sec:exp_results}
We first demonstrate how \sys{} supports different batch sizes by evaluating \textbf{NN5} on \textbf{Scenario 1} (Section~\ref{sec:elastic_batch}). We also compare \sys{} with prior work on private PaaS as \textbf{NN5} is the predominantly used benchmark. Then, in Section~\ref{sec:scalability}, we evaluate \textbf{NN20} on \textbf{Scenario 3} to discuss \sys's scalability aspects with the number of model-providers. Finally, we demonstrate \sys{}'s application and model agility by evaluating the more complex model \textbf{NN50} in all scenarios of Section~\ref{sec:design} (Section~\ref{sec:model_agility}).

\subsubsection{Elastic Data Packing.}\label{sec:elastic_batch}
\begin{figure}[t]
	\centering
	\begin{subfigure}[t]{0.49\textwidth}
		\centering
		\includegraphics[width=\textwidth]
		{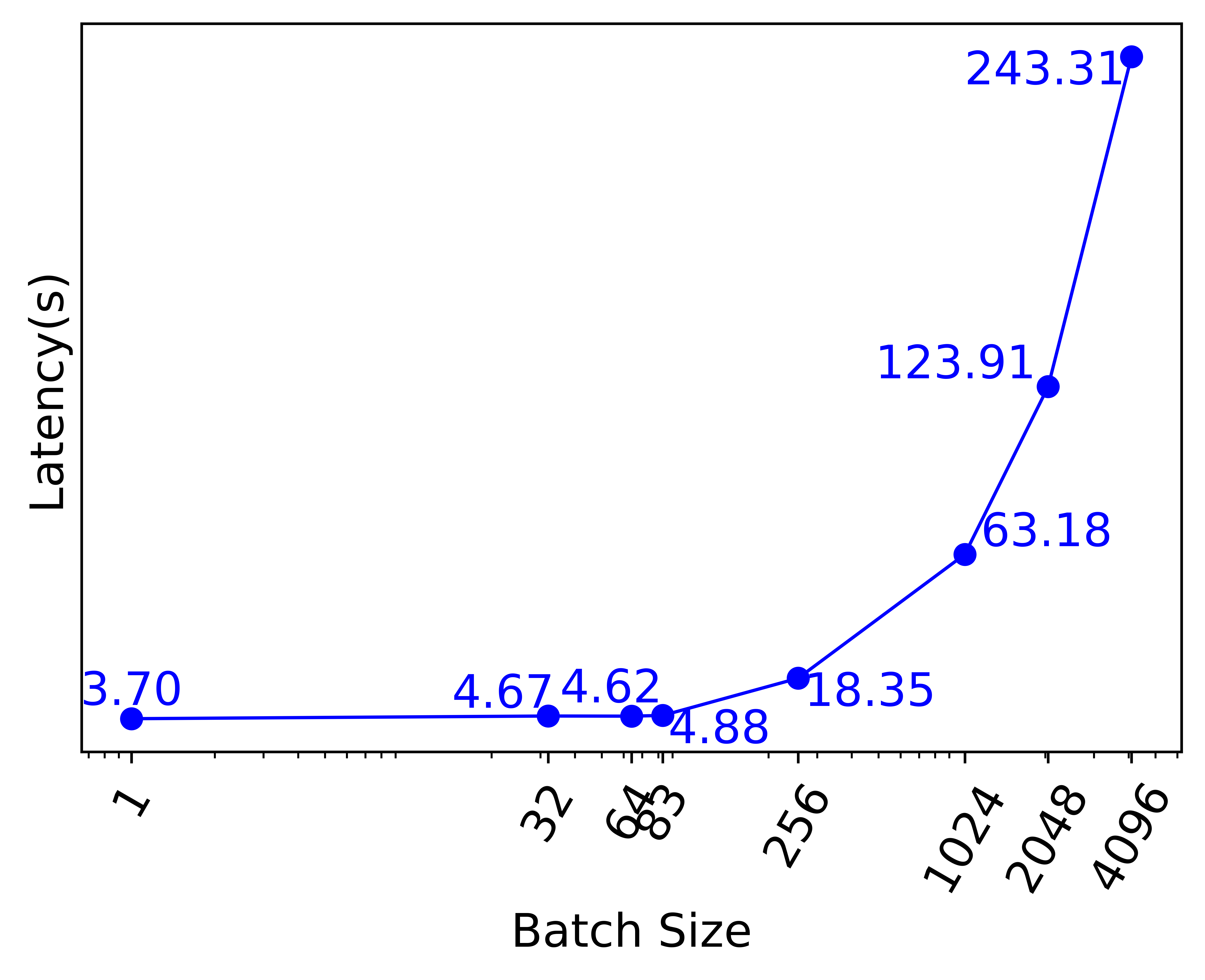}
		\caption{\sys{}'s latency for \textbf{NN5} and different batch sizes.}
		\label{fig:latency}
	\end{subfigure}
    \hfill
	\begin{subfigure}[t]{0.49\textwidth}
		\centering
		\includegraphics[width=\textwidth]
		{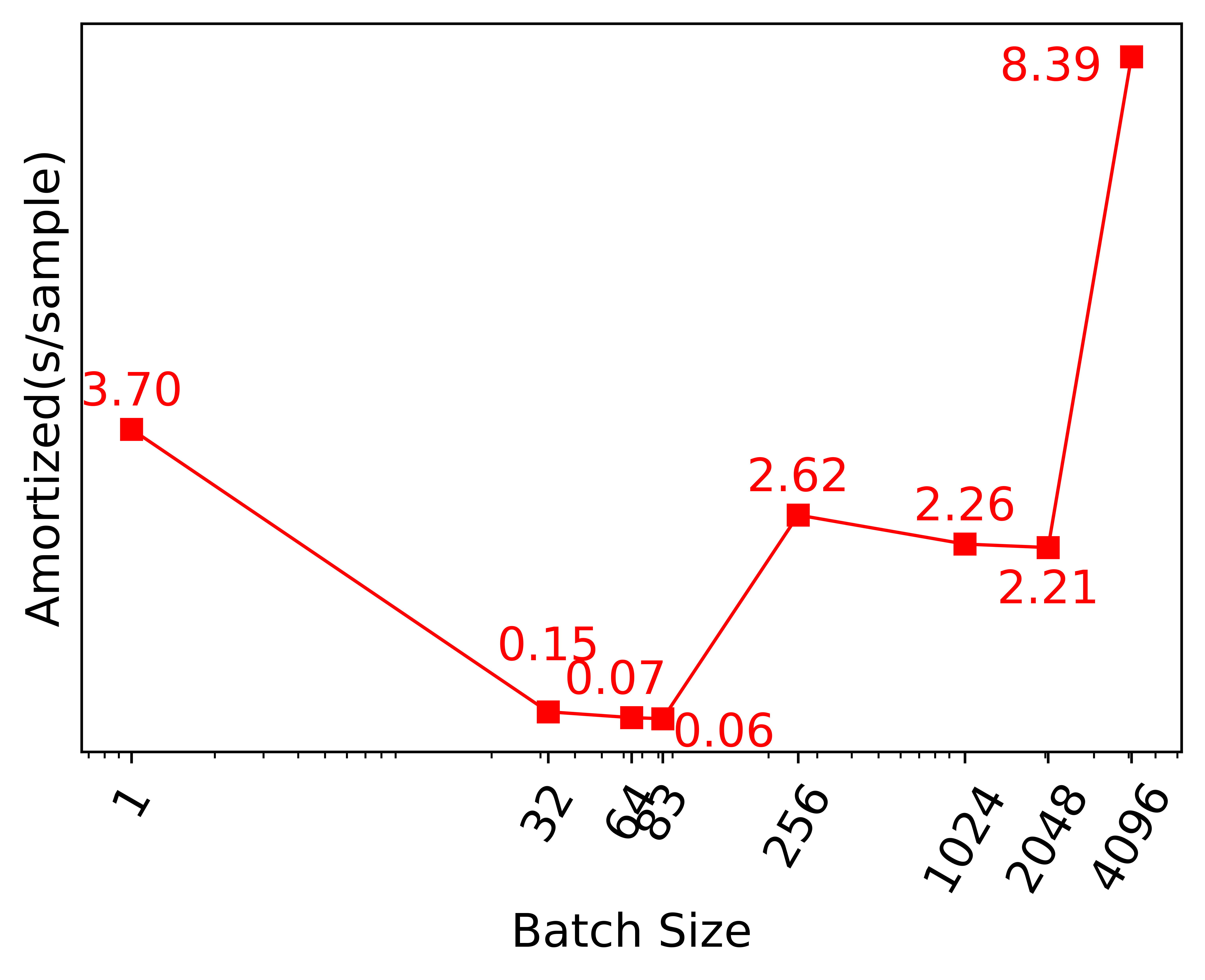}
		\caption{\sys{}'s amortized runtime for \textbf{NN5} and different batch sizes.}
		\label{fig:amortized}
	\end{subfigure}
	\caption{\sys{}'s amortized runtime with different batch sizes.}
 \label{fig:batch}
\end{figure}

We demonstrate the benefits of our packing approach (Section~\ref{sub:packing}), by benchmarking \textbf{NN5}~\cite{CryptoNets} in the traditional PaaS setting (\textbf{Scenario 1}) for various batch sizes. For this experiment, \sys{} heuristically estimates the optimal batch size for throughput at 83, as described in Section~\ref{sub:optimizations}; this is experimentally confirmed by observing Figures~\ref{fig:latency} and~\ref{fig:amortized}. In particular, Figure~\ref{fig:latency} shows \sys{}'s latency for varying batch sizes up to $4,096$ in semi-log scale. We observe a linear increase in latency after the optimal size. This is expected, as \sys{} automatically splits batch sizes larger than the optimal size into sub-batches of optimal size, and processes them sequentially. Figure~\ref{fig:amortized} shows the amortized runtime of \sys{} for variable batch sizes: We observe that a batch of size $83$ is indeed the optimal point which minimizes the amortized runtime (or maximizes the throughput). Finally, we compare \sys{}'s performance with related works that evaluate \textbf{NN5} in the same application scenario with polynomial activation functions (thus, we exclude Gazelle~\cite{Gazelle} which relies on Garbled Circuits). Table~\ref{tab:cn_bench} shows that \sys{}'s performance is on par with or better than previous works, while providing enhanced flexibility in terms of batch size. The approach followed by CryptoNets and inspired works \cite{CryptoNets,fasterCN,boemer2019ngraph} allows them to achieve a good throughput by processing large batches of data items (up to $\mathcal{N}/2$), but their runtime is independent of the batch size (hence, it will not decrease for smaller batches as per Table~\ref{tab:cn_bench}). Conversely, the approach followed by LoLa \cite{brutzkus2019low} achieves low latency for a single sample, but cannot amortize the runtime when processing multiple samples. With \sys{}, the end-user can define its custom batch size without a major impact on performance.

\begin{table}[h]
\centering
\setlength{\tabcolsep}{2.9pt}
\begin{tabular}{cccc}
\toprule
                        & \multicolumn{3}{c}{\textbf{Latency (s)}}             \\
                        \toprule
Framework          & Batch size = 1 & Batch size = 83 & Batch size = 4,096 \\
\midrule
CryptoNets~\cite{CryptoNets}              & \textcolor{black}{250}             & \textcolor{black}{250}             & \textcolor{black}{250}               \\
Faster CryptoNets~\cite{fasterCN}       & \textcolor{ black}{39.1}           & \textcolor{black}{3,245}            & \textcolor{black}{160,153 }                \\
LoLa~\cite{brutzkus2019low}                    & \textcolor{black}{2.2}           & \textcolor{black}{182.6 }             & \textcolor{black}{8,951 }                \\
nGraph-HE2~\cite{boemer2019ngraph}\footnotemark         & \textcolor{black}{2.05   }   & \textcolor{black}{2.05 }           & \textcolor{black}{2.05}               \\
\midrule
\sys{} & {3.7}            &     {4.08 }        & {243.4 }          \\ \bottomrule     
\end{tabular}
\caption{Latency comparison between \sys{} and prior encrypted frameworks for the evaluation of \textbf{NN5} and various batch sizes.}
\label{tab:cn_bench}
\end{table}

\begin{figure}[h]
    \centering
    \begin{minipage}{0.49\textwidth}
        \centering
        \begin{table}[H]
        \centering
        \begin{tabular}{ccc}
        \toprule
          \textbf{\begin{tabular}[c]{@{}c@{}}\# of Parties\end{tabular}} &
          \textbf{\begin{tabular}[c]{@{}c@{}}Latency\\ (s)\end{tabular}} &
          \textbf{\begin{tabular}[c]{@{}c@{}}Throughput\\ (samples/s)\end{tabular}} \\
        \toprule
        \textbf{\begin{tabular}[c]{@{}c@{}}3\end{tabular}} & 245.58 ($\pm 0.50$) & 1.19 \\
        \midrule
        \textbf{\begin{tabular}[c]{@{}c@{}}5\end{tabular}} & 238.15 ($\pm 4.12$) & 1.22 \\
        \midrule
        \textbf{\begin{tabular}[c]{@{}c@{}}10\end{tabular}} & 278.19 ($\pm 9.11$) & 1.05\\
        \midrule
        \textbf{\begin{tabular}[c]{@{}c@{}}20\end{tabular}} & 354.17 ($\pm 10.66$) & 0.82\\
        \bottomrule
        \end{tabular}%
        \caption{\sys{}'s performance for \textbf{NN20} on \textbf{Scenario 3} (Section~\ref{sub:scenario3}) with increasing number of parties (model-providers).}
        \label{tab:nn20}
        \end{table}
    \end{minipage}
    \hfill
    \begin{minipage}{0.49\textwidth}
        \centering
        \includegraphics[width=0.8\textwidth]{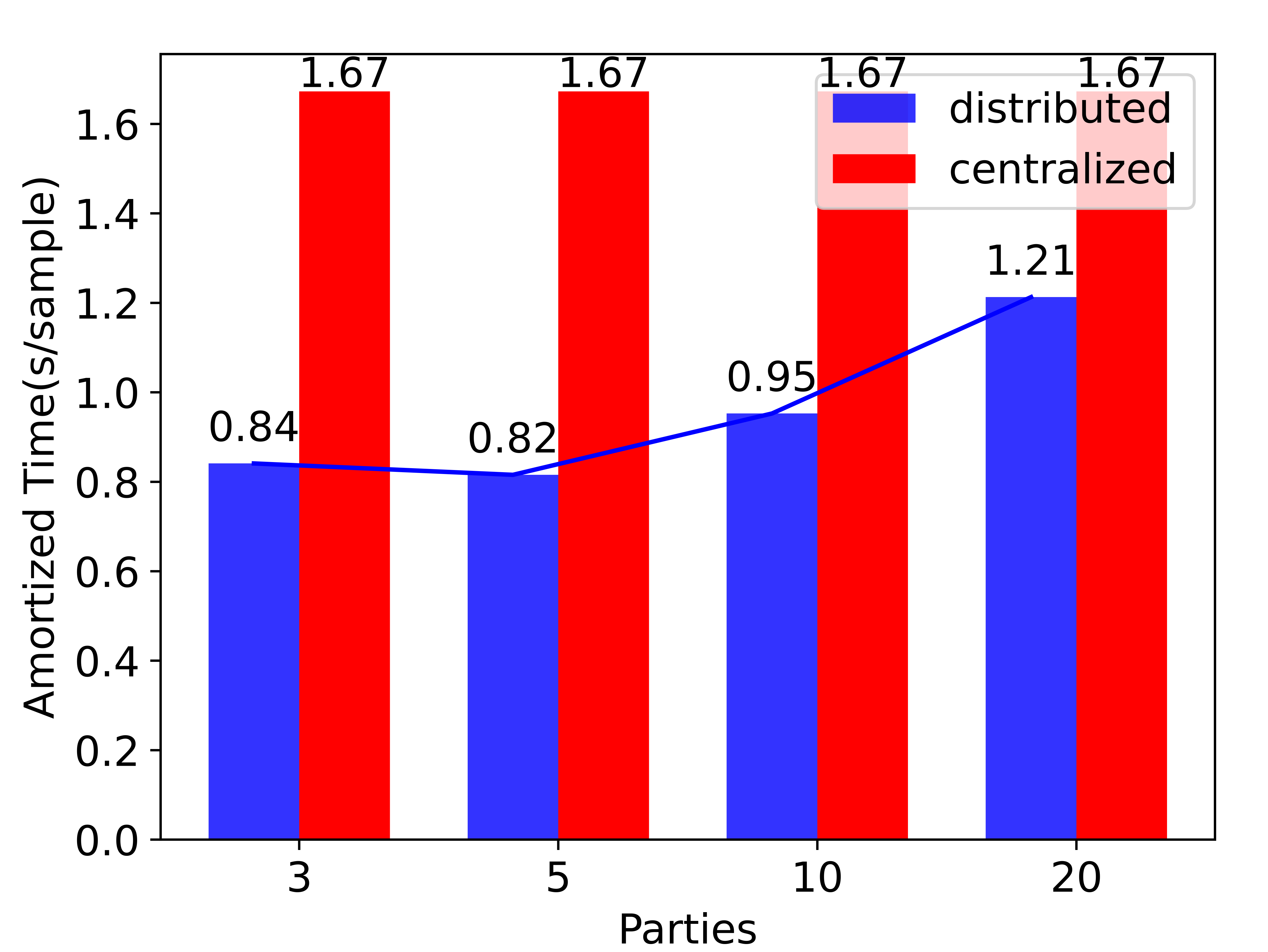}
        \caption{Benchmarking decentralized vs. centralized bootstrapping on encrypted \textbf{NN20} for variable number of parties (model-providers).}
        \label{fig:distr_vs_centr_btp}
    \end{minipage}
\end{figure}

\footnotetext{While \sys{} and related works \cite{CryptoNets,fasterCN,brutzkus2019low} employ similar hardware for testing, we note that nGraph-HE2~\cite{boemer2019ngraph} employs compiler optimizations and a more performant hardware with $376$GB of RAM and $112$ cores.}

\subsubsection{Interactive MPC Protocols.}\label{sec:scalability}
We evaluate \textbf{NN20} in \textbf{Scenario 3} where the model is trained and retained under encryption by multiple parties using a privacy-preserving collaborative training framework~\cite{poseidon,spindle,rhode} (Section~\ref{sub:scenario3}). Note that this scenario requires the use of the collective bootstrapping protocol $\textsf{CBootstrap}(\cdot)$, thus, it is not supported by prior encrypted inference frameworks. Table~\ref{tab:nn20} shows \sys{}'s latency and throughput for increasing number of parties, while in Figure~\ref{fig:distr_vs_centr_btp} we compare \sys{}'s amortized runtime when employing $\textsf{CBootstrap}(\cdot)$ versus the centralized bootstrapping. Overall, we observe a linear increase (decrease) in \sys{}'s latency (throughput) as the number of parties increases. We note that the $\textsf{CBootstrap}(\cdot)$ operation is executed in an asynchronous fashion by the master model provider, i.e., the protocol is initiated concurrently with all the model providers and the output is generated as soon as the last party provides its share. For this reason, we can even experience lower latency when increasing the number of parties by a limited amount ($3$ vs. $5$), as the protocol becomes particularly sensible to the network conditions. In any case, the benefits of employing $\textsf{CBootstrap}(\cdot)$ over the centralized version (when possible) are evident, as the former enables refreshing the ciphertext noise with an efficient interactive protocol, rather than with a computationally expensive homomorphic circuit.

\subsubsection{Application and Model Agility.}\label{sec:model_agility}


Finally, we demonstrate the high degree of flexibility offered by \sys{}, both in terms of variety of enabled use-cases  and supported architectures, by evaluating a more complex model on all the scenarios described in Section~\ref{sec:design}. In particular, we benchmark \sys{} with \textbf{NN50} and a batch of $585$ samples on: (i) \textbf{Scenario 1} with encrypted data and a plaintext model (Section~\ref{sub:scenario1}), (ii) \textbf{Scenario 2} with an encrypted model and plaintext data (Section~\ref{sub:scenario2}), and (iii) \textbf{Scenario 3} where the encrypted model is kept by $N{=}3$ model-providers and encrypted data. Note that evaluating \textbf{NN50} in \textbf{Scenarios 1} and \textbf{2} requires the invocation of the centralized bootstrapping operation, that is not supported by most of the related works~\cite{CryptoNets,Gazelle,boemer2019ngraph,brutzkus2019low}. 

Table~\ref{tab:nn50} shows the performance results for all scenarios. First, we note that by leveraging on our data packing approach and processing multiple samples in a SIMD fashion, \sys{} achieves reasonable runtime given the complexity of the \textbf{NN50} model (\textbf{Scenario 1}). For reference, the original work by Chillotti et al. achieves at best an amortized time of $37.69$s/sample and a throughput of $0.02$samples/s. Then, we also observe that \sys{}'s generic optimizations enable the efficient evaluation of encrypted models: Evaluating \textbf{NN50} under encryption on \textbf{Scenario 2}, which involves a matrix multiplication, addition, polynomial activation, and centralized bootstrapping operations, is only $\sim$7$\%$ slower than evaluating a plaintext model evaluation (c.f. \textbf{Scenario 1}). \sys{} achieves the best performance results on \textbf{Scenario 3} thanks to its support for interactive multiparty protocols such as collective bootstrapping (Section~\ref{sec:scalability}). Overall, we remark that \sys{} is the first framework for encrypted inference that can support all these application scenarios.


\begin{table}[t]
\centering
\begin{tabular}{ccccc}
\toprule
 &
  \textbf{Latency(s)} &
  \textbf{\begin{tabular}[c]{@{}c@{}}Amortized\\ (s/sample)\end{tabular}} &
  \textbf{\begin{tabular}[c]{@{}c@{}}Throughput\\ (samples/s)\end{tabular}} &
  \textbf{\begin{tabular}[c]{@{}c@{}}Avg. latency/layer\\(s)\end{tabular}} \\
\toprule
\textbf{\begin{tabular}[c]{@{}c@{}}Plaintext model\\ Encrypted data\\ (Scenario 1)\end{tabular}} & 2,496.83 & 4.26 & 0.234  & 48.95 \\
\midrule
\textbf{\begin{tabular}[c]{@{}c@{}}Encrypted model\\ Plaintext data\\ (Scenario 2)\end{tabular}} & 2,699.75 & 4.62 & 0.216  & 52.93 \\
\midrule
\textbf{\begin{tabular}[c]{@{}c@{}}Encrypted model\\ Encrypted data\\ (Scenario 3)\end{tabular}} & 613.52  & 2.09  & 0.476 & 12.02\\
\bottomrule
\end{tabular}%
\caption{\sys{}'s performance for \textbf{NN50} in \textbf{Scenarios 1}, \textbf{2}, and \textbf{3}  (Section~\ref{sec:design}). For \textbf{Scenario 3}, the number of model-providers is $N{=}3$.}
\label{tab:nn50}
\end{table}



\section{Conclusion}
In this work, we presented \sys{}, an agile framework for privacy-preserving deep neural network inference using homomorphic encryption. Thanks to our hybrid design that leverages on HE and its multiparty variant, and generic setting-agnostic optimizations, \sys{} can support various and novel scenarios for encrypted inference featuring untrusted model providers and clients. These scenarios include: (i) the client sending encrypted data to an untrusted model-provider for inference, (ii) the model-provider sending an encrypted model to a client for local inference (without the need of mutual trust between them), and (iii) the client sending the encrypted data to a cohort of model-providers holding an encrypted model. Thus, \sys{} extends the applicability of privacy-preserving PaaS beyond previous works. Moreover, with our intuitive and flexible input data packing scheme, \sys{} can be adapted to various deep neural network architectures and can accommodate diverse application requirements, being able to process an arbitrary number of samples without incurring major performance loss. Our experimental results show that the simplicity of our packing approach and the agility of our framework does not harm its performance as it is on par with, and occasionally better than, state-of-the-art related works, while introducing an increased degree of flexibility over previous works.


%
%
\bibliographystyle{splncs04}
\bibliography{bibfile}
\end{document}